\documentclass[]{spie}  %>>> use for US letter paper

\usepackage{amsmath,amsfonts,amssymb}
\usepackage{graphicx}
\usepackage[colorlinks=true, allcolors=blue]{hyperref}
\usepackage{subfig}
\usepackage{tabularx}
\title{Adversarial Threat Vectors and Risk Mitigation for Retrieval-Augmented Generation Systems}
\author{Chris M. Ward, Josh Harguess\\Fire Mountain Labs\\
San Diego, CA, USA \\
\{chris, harguess\}@firemountainlabs.com}

\begin{document}
\maketitle

\begin{abstract}

Retrieval-Augmented Generation (RAG) systems, which integrate Large Language Models (LLMs) with external knowledge sources, are vulnerable to a range of adversarial attack vectors. This paper examines the importance of RAG systems through recent industry adoption trends and identifies the prominent attack vectors for RAG: prompt injection, data poisoning, and adversarial query manipulation. We analyze these threats under risk management lens, and propose robust prioritized control list that includes risk-mitigating actions like input validation, adversarial training, and real-time monitoring.
\end{abstract}

\section{Introduction}
\label{sec:intro}

Retrieval-Augmented Generation (RAG) systems extend the capabilities of Large Language Models (LLMs) by incorporating real-time, external data sources to enhance response relevance and accuracy. Since their introduction in 2020~\cite{lewis2020retrieval}, adoption has surged; recent reports indicate enterprise use exceeded 50\% in 2024, up from 31\% the prior year~\cite{menlo2024}. RAG systems are increasingly embedded in critical industries such as finance, healthcare, and legal services, creating new challenges for AI security~\cite{babeanu2025}.

While RAG systems deliver flexible and up-to-date outputs, their reliance on external, mutable data introduces unique security risks.\cite{xue2024badrag} This creates a fundamental tension between functionality and security, where protective measures must safeguard system integrity without unduly restricting utility~\cite{rieder2021mapping,Dahj2022MasteringCI,ai2023artificial,Shostack2014}.

In Sections \ref{sec:llm} and \ref{sec:rag_background}, we provide a brief technical background on LLMs and RAG system architecture. We discuss the AI Security Pyramid of Pain\cite{ward2024ai}, a structured framework for ranking controls/ mitigations by robustness, in Section \ref{sec:pyramid_intro}. Section \ref{sec:cwe_background} gives an overview of the MITRE Common Weakness Enumeration (CWE) framework and its application to AI systems, distinguishing between system weaknesses and vulnerabilities.

Section \ref{sec:analysis} of this paper provides a detailed analysis centered on a structured threat modeling process process applied to a generic Retrieval-Augmented Generation (RAG) system. The methodology unfolds in several key stages, commencing with the definition of the system's scope and objectives in Section \ref{sec:define_scope}, followed by a thorough decomposition of its architecture to identify critical components and data flows in Section \ref{sec:decompose_system}. Building on this foundation, Section \ref{sec:identify_threats} identifies and examines significant risks to an operation RAG architecture, such as sensitive information disclosure and RAG system poisoning, referencing frameworks like MITRE ATLAS and the OWASP Top 10 for LLM Applications. We assess and prioritize these identified risks, including a quantification of inherent risk, in Section \ref{sec:risk_quantification}. We discuss risk mitigation controls, and their prioritization using the AI Security Pyramid of Pain to maximize adversary disruption in Sections \ref{sec:risk_controls} and \ref{sec:risk_control_prioritization} respectively. We examine remaining residual risk in Section \ref{sec:validation_residual_risk} and validate the effectiveness of our mitigation strategy. We discuss key findings and propose areas for future work in Section \ref{sec:discussion_future_work}. We conclude our analysis in Section \ref{sec:conclusion}.

\section{Background}
\label{sec:background} 
We begin by exploring Large Language Models (LLMs), as they are central to our methodology and findings, covering their architectural principles, training paradigms, and overall impact.
\subsection{Large Language Models}
\label{sec:llm} 

In recent years, Large Language Models (LLMs) have emerged as a dominant paradigm in natural language processing, built upon the transformer architecture introduced by Vaswani et al. \cite{vaswani2017attention}. These models implement a self-attention mechanism that enables parallel processing of sequential data while maintaining awareness of contextual relationships between tokens (words in sentence, or subsections of an image). Modern LLMs typically employ decoder-only architectures with autoregressive training objectives, optimizing next-token prediction across massive text corpora\cite{zhao2023survey}.
The computational backbone of LLMs consists of multi-head self-attention layers alternating with feed-forward neural networks. Each attention head computes query, key, and value projections to model token interactions across variable distances. This architecture enables the capture of complex linguistic patterns, including long-range dependencies, syntactic structures, and semantic relationships.

\subsection{Retrieval-Augmented Generation (RAG)}
\label{sec:rag_background} 
RAG systems are designed to enhance the capabilities of large language models (LLMs) by integrating real-time data retrieval mechanisms. This allows RAG systems to generate responses that are not only based on pre-trained knowledge but also enriched with up-to-date, context-specific information retrieved from external sources. The RAG architecture, as illustrated in Figure \ref{fig:rag}, typically comprises several main conceptual components:

\begin{figure}[ht]
    \centering
    \includegraphics[width=\textwidth]{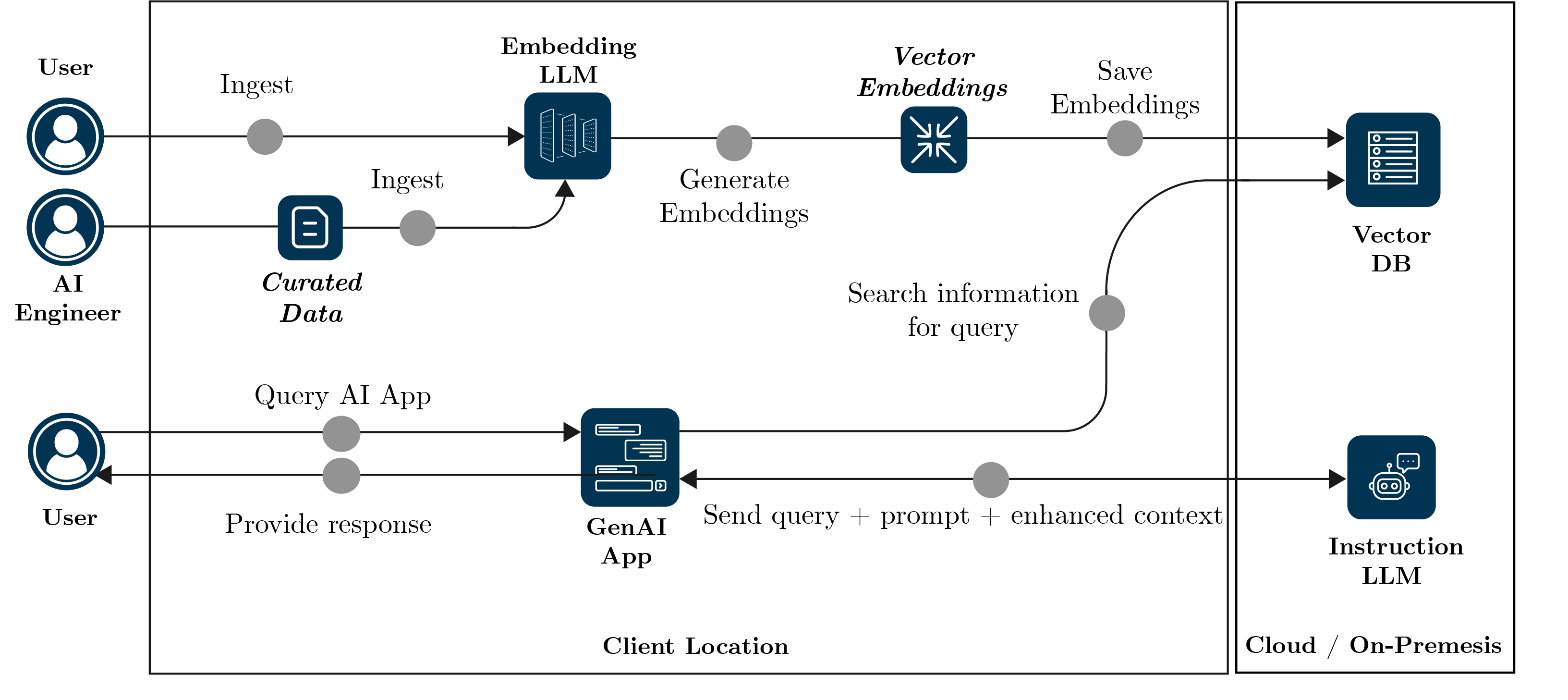}
    \caption{A generalized Retrieval Augmented Generation (RAG) Architecture}
    \label{fig:rag}
\end{figure}

\subparagraph{Retrieval Component:}
The retrieval phase fetches relevant information from external sources such as structured databases, web data, internal documents, or specialized repositories. 

\textit{Example}: When a user asks a technical question, the system retrieves internal design documents or API specifications as supporting context.

\subparagraph{Generative Component:}
The generative model, typically a large language model, combines the user query with the retrieved context to produce an accurate, contextually relevant response.
    
\textit{Example:} Given a user query and retrieved code documentation, the model generates a natural language explanation of how a function works.

\subparagraph{Embedding Model:}
Embedding LLMs convert text into compact vector representations that compress semantic information. 

\textit{Example:} A transformer-based embedding model transforms an academic paper's abstract into a vector, enabling efficient clustering and retrieval of similar research.

RAG architecture improves on traditional language models by integrating live data retrieval with text generation. This yields more current and accurate answers, particularly for time-sensitive queries. It can also be tailored with domain-specific knowledge for areas such as customer support, healthcare, finance, and research. By accessing up-to-date information, the system stays relevant without the need for resource-intensive model training or fine-tuning.

While RAG systems offer powerful advantages, their dependence on external data sources also introduces unique security challenges. The retrieval process, in particular, can expose the system to vulnerabilities if the external data is tampered with or contains malicious content. Understanding this AI architecture as a holistic system is crucial for identifying and mitigating these risks.

\subsection{AI Security Pyramid of Pain}
\label{sec:pyramid_intro} 

\begin{figure}[ht]
    \centering
    \includegraphics[width=0.6\textwidth]{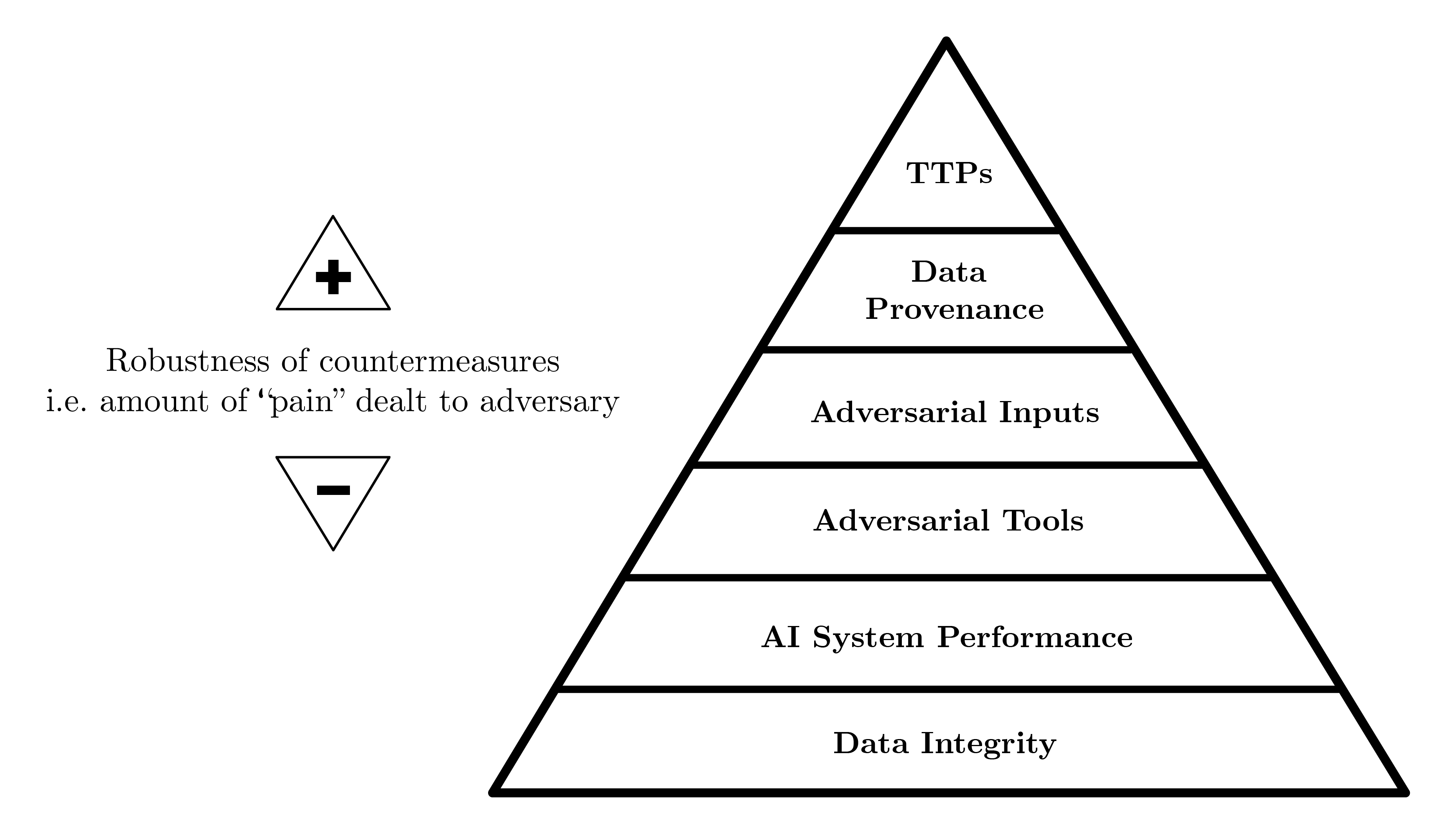}
    \caption{The AI Security Pyramid of Pain (Ward et. al.\cite{ward2024ai})}
    \label{fig:pyramid_of_pain}
\end{figure}

The AI Security Pyramid of Pain\cite{ward2024ai} (shown in Figure \ref{fig:pyramid_of_pain}) is a structured framework for categorizing and prioritizing adversarial countermeasures for AI-enabled systems. By providing a hierarchical approach to driving down risk, the framework enables organizations to systematically address weaknesses and vulnerabilities across multiple layers, from foundational data integrity controls to undermining the adversaries' tactical playbook.

We apply the AI Security Pyramid of Pain to our generic RAG system in order to enhance operational resilience by prioritizing more robust defenses. This approach not only addresses known weaknesses, but also anticipates an evolving adversarial landscape, resulting in an adaptive security strategy for our RAG system.

\subsection{Common Weaknesses}
\label{sec:cwe_background}

The CWE catalog helps identify and classify weaknesses in software, hardware, and other digital systems. Created in 2006, CWE provides a standardized way to describe security flaws so that developers, businesses, and researchers can work toward fixing them \cite{mitre_cwe_overview}.
In recent years, CWE has expanded its focus to include AI-related weaknesses \cite{cwe_ai_wg}.

Because configuration management is complex for AI systems, we advocate for the use of the MITRE Common Weakness Enumeration (CWE) framework, which is more generalizable and better suited than vulnerability mapping (CVSS, etc) in the AI domain.\cite{mitre2006cwe, martin2008common}

The CWE framework prefers the term \textit{``weakness"} over \textit{``vulnerability"} because it focuses on \textit{potential} security flaws that could lead to vulnerabilities if not addressed. A weakness represents an underlying issue in software, AI models, or hardware, whereas a vulnerability refers to an \textit{actively exploitable} instance of a weakness. Table~\ref{tab:weakness_vs_vulnerability} highlights the core distinctions between these two terms with examples.

\begin{table}[h]
    \caption{Comparison of Weakness vs. Vulnerability}
    \label{tab:weakness_vs_vulnerability}
    \centering
    \begin{tabular}{|l|p{5cm}|p{5cm}|}
        \hline
        \textbf{Term} & \textbf{Definition} & \textbf{Example} \\ \hline
        \textbf{Weakness} & A flaw, mistake, or security oversight in design, code, or system logic that could potentially be exploited. & A web application accepts user input without proper validation. (CWE-20: Improper Input Validation\cite{CWE-20}) \\ \hline
        \textbf{Vulnerability} & A confirmed instance where a weakness has been exploited or has led to a security breach. & An attacker injects malicious SQL commands due to lack of input validation, leading to data leaks. (CVE-2023-5423: SQL Injection in Online Pizza Ordering System\cite{cve-2023-5423}) \\ \hline
    \end{tabular}
\end{table}

CWE provides a pathway for cataloging potential systemic issues before they become exploitable, recognizing that not all weaknesses immediately manifest as security vulnerabilities but can still compromise system reliability, performance, and information integrity. This approach provides a broader coverage than traditional vulnerability assessments. Table \ref{tab:ai_weaknesses} provides two concrete examples of AI weaknesses covered by the CWE program.

\begin{table}[h]
  \caption{Examples of AI Weaknesses and Their Potential Exploits}
    \label{tab:ai_weaknesses}
    \centering
    \begin{tabular}{|p{5cm}|p{5cm}|p{5cm}|}
        \hline
        \textbf{Weakness (CWE ID)} & \textbf{Description} & \textbf{Potential Vulnerability} \\ \hline
        
        CWE-1039: Inadequate Handling of Adversarial Input\cite{CWE-1039} & AI fails to detect subtle manipulations in input data. & A facial recognition system is tricked into granting unauthorized access using an altered image. \\ \hline
        CWE-1426: Improper Validation of AI Output\cite{CWE-1426} & AI-generated text is not properly checked, leading to misinformation or bias. & Attackers manipulate AI responses to spread false news or harmful content. \\ \hline
    \end{tabular}
\end{table}

By focusing on weaknesses, CWE helps organizations improve system security before vulnerabilities emerge. This is especially important in AI, where potential weaknesses like incorrect outputs\cite{CWE-1426}, or inadequate handling of adversarial inputs\cite{CWE-1039}, can lead to serious security and ethical concerns if not addressed early.

\section{Analysis}
\label{sec:analysis}

\begin{figure}
    \centering
    \includegraphics[width=0.96\textwidth]{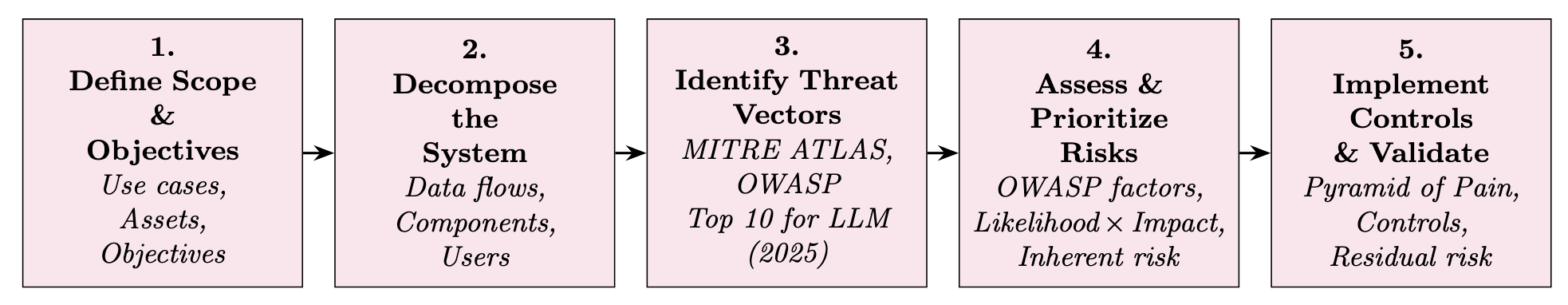}
\vspace{.5cm}
\caption{A generalized threat modeling process used in our analysis.}
\label{fig:threat-modeling-process}
\end{figure}

Our analysis applies a generalized threat modeling process that can be applied to AI-enabled systems. Inspired by Microsoft's Security Development Lifecycle (SDL) threat modeling methodology~\cite{microsoftThreatModeling} and prior work in Doyle, et al.\cite{doyle2021vulnerability}, this approach decomposes the system, identifies threats, and prioritizes them based on real-world likelihood and impact. This process, shown in Figure~\ref{fig:threat-modeling-process}, enables the identification of high-impact, system-level threats; it supports actionable mitigation strategies tailored to the dynamics of generative AI.

\clearpage

We conduct five sequential stages of analysis: (1) define scope and objectives; (2) decompose system architecture and data flows; (3) identify threat vectors using MITRE ATLAS and OWASP Top 10 for LLMs; (4) assess and prioritize risks based on OWASP factors, and inherent risk (likelihood × impact), ; and (5) implement controls, validate, and measure residual risk. The process can be repeated until risks are drawn-down to acceptable levels. 

\subsection{Define the Scope and Objectives}
\label{sec:define_scope}
Our simple RAG is designed to integrate natural language understanding with access to an enterprise document corpus. As shown in Figure~\ref{fig:rag}, the architecture accepts user queries via a chat interface, retrieves semantically relevant documents from a vector database, and injects both the prompt and retrieved context into the LLM's input. The system components include embedding generation pipelines, retrieval APIs, document ingestion workflows, and the LLM inference layer - each introducing unique risks related to data exposure, integrity, and model behavior.

The business use case for the RAG system is enterprise knowledge management. It supports employees by providing natural-language access to internal documentation such as policies, procedures, product manuals, and historical records. By replacing keyword-based search with contextualized responses, the system improves employee efficiency, reduces support latency, and ensures consistent access to authoritative information across business units.

\subsection{Decompose the System}
\label{sec:decompose_system}
To understand where attack vectors may emerge in a system, we decompose the RAG system into its functional components and analyze how data flows through each. Each component changes the attack surfaces, introducing new vectors like document ingestion, embedding generation, prompt construction, and LLM inference.

The attack surface expands across the ingestion of untrusted content, construction of retrieval indexes, serialization of vector representations, and retrieval logic. Additionally, the LLM interface itself becomes a target due to its capability to interpret and act upon adversarially crafted inputs.

\begin{figure}[ht]
%\includesvg[width=0.95\textwidth]{figures/RAG Attack Surface.svg}

\includegraphics[width=0.95\textwidth]{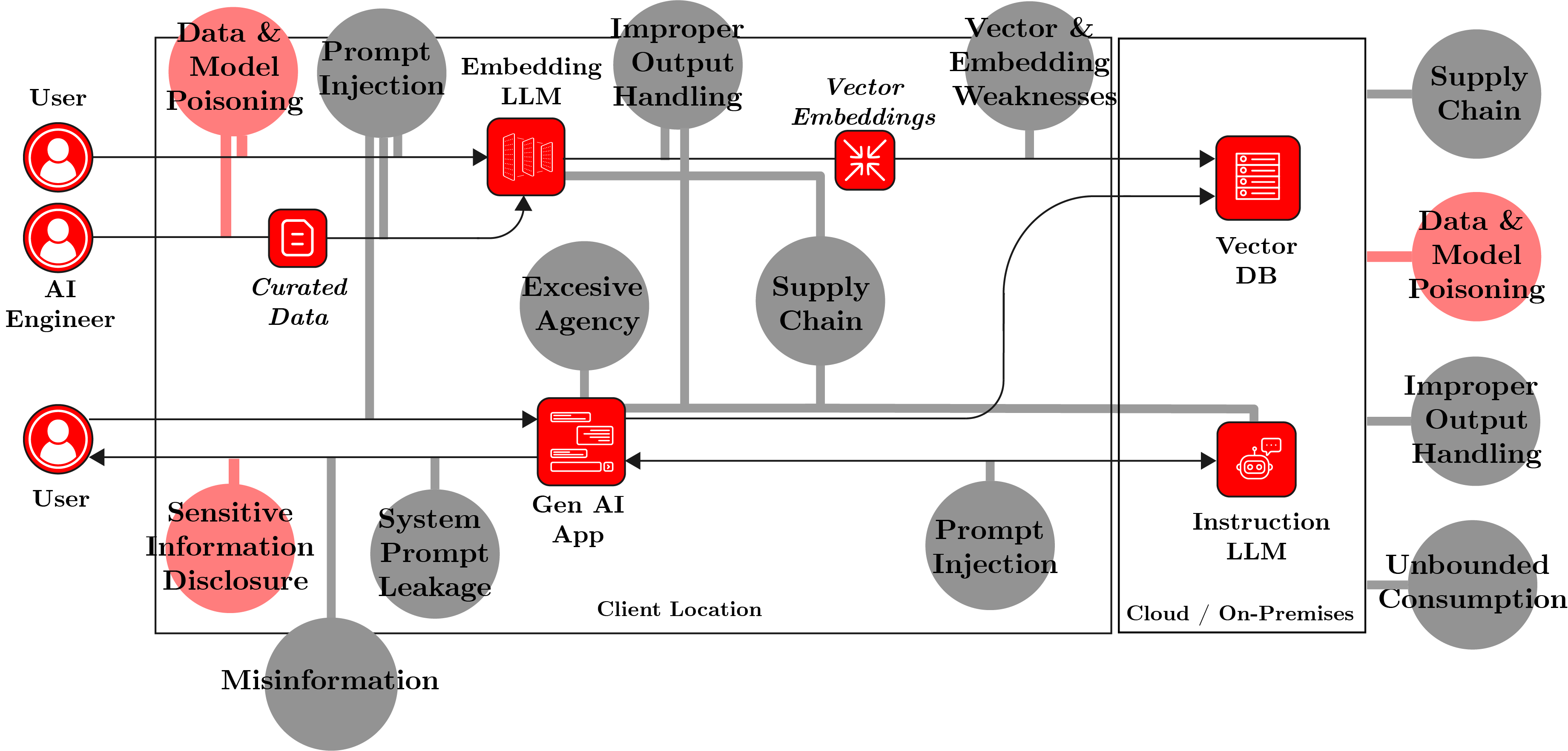}
\vspace{0.20cm}
\caption{RAG attack surface overlay. Approximate entry points for common weaknesses are shown.}
\label{fig:attack-surface}
\end{figure}

Lacking a pre-existing architecture diagram, one may opt to construct a high-level visual representation, document the system's components and data flows, or infer the architectural design by analyzing available technical artifacts, source code, and Open source intelligence (OSINT). Figure~\ref{fig:attack-surface} illustrates the attack surface overlaid on the system's logical architecture. This diagram highlights potential adversary access points, such as document upload pipelines, embedding APIs, vector store queries, prompt injection vectors, and inference endpoints.

\subsection{Identify Threat Vectors}
\label{sec:identify_threats}
This section first introduces two primary threat models and then maps them to the relevant risk layers within a RAG system. The objective is to provide a clear understanding of how specific adversarial techniques align with vulnerabilities across distinct operational domains. In Figure \ref{fig:attack-surface}, we map the comprehensive RAG attack surface, detailing how adversaries can leverage weaknesses and vulnerabilities across ingestion, retrieval, and generation components to compromise system integrity. While a broader range of threat models could (and should) be considered, this focus allows for a thorough examination pertinent to the core objectives and constraints of this work.

\subsubsection{Threat Model I: Sensitive Information Disclosure in RAG Systems}

This threat model focuses on adversarial efforts to extract or expose confidential information from RAG systems. These systems, which combine LLMs with external knowledge retrieval components, are vulnerable to prompt-based manipulation. Attackers often employ direct or indirect prompt injection techniques to manipulate retrieval queries, resulting in the unintended disclosure of sensitive content.

Sensitive information disclosure typically involves the exposure of Personally Identifiable Information (PII), financial records, proprietary algorithms, internal business logic, or system-level prompts. Vulnerabilities in prompt handling and insufficient guardrails around retrieved content allow LLMs to surface confidential information not intended for output.

Key adversarial techniques include:

\begin{itemize}
    \item \textbf{Membership inference attacks} to identify whether specific records were used during model training~\cite{mitre:AML.T0024.000}.
    \item \textbf{Model inversion attacks} that reconstruct training data from model outputs~\cite{mitre:AML.T0024.001}.
    \item \textbf{System Prompt leakage}, where system or retrieval prompts containing confidential information are exposed to the end user~\cite{mitre:AML.T0051.000, OWASP_LLM07_PromptLeakage}.
    \item \textbf{Embedding exploitation}, where attackers manipulate or query vector stores to extract hidden data~\cite{mitre:AML.T0024.002, OWASP_LLM08_VectorEmbedding}.
\end{itemize}

\begin{figure}[ht]
    \centering

    % First subfigure (Image)
    \subfloat[RAG Architecture labeled with Tactics, Techniques, and Procedures (TTPs) for Sensitive Information Disclosure Attack\label{fig:info_leak}]{
        \includegraphics[width=0.90\textwidth]{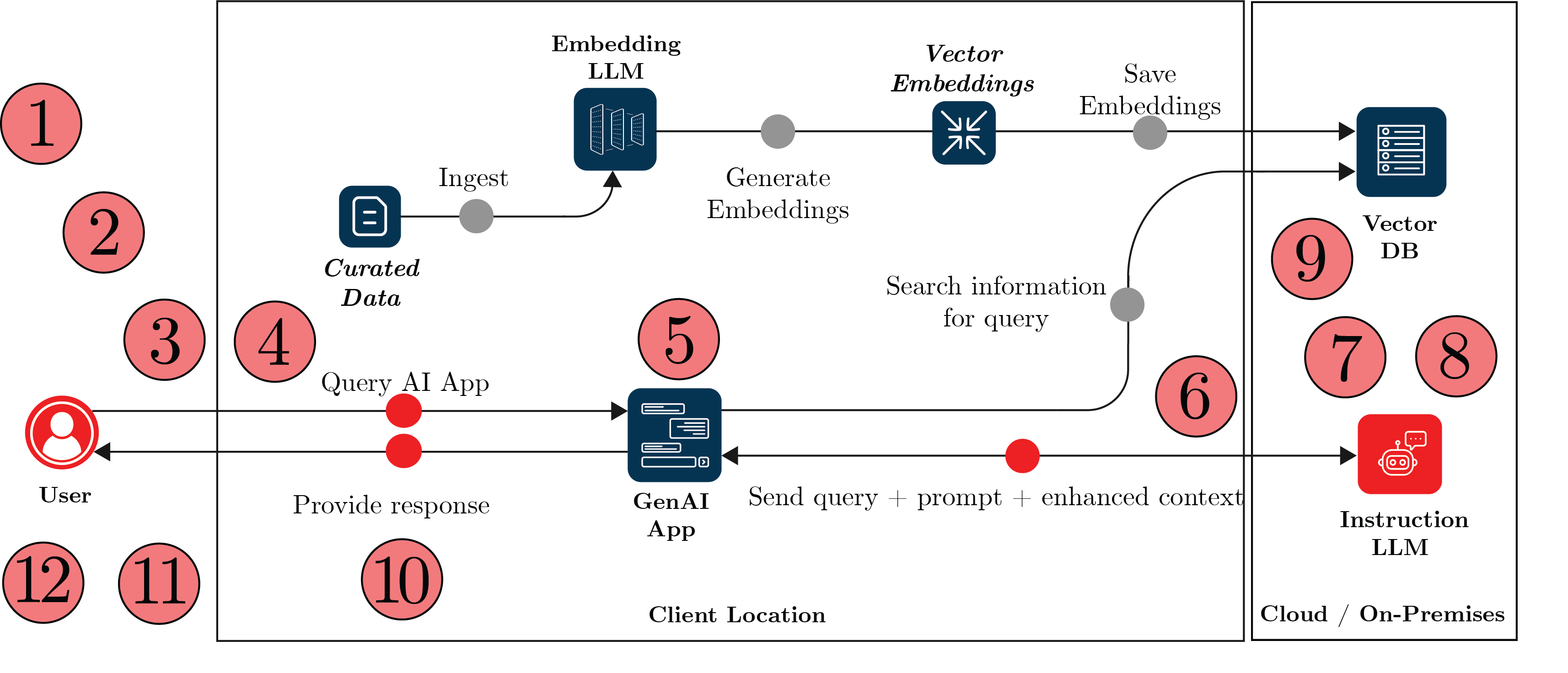}
    }\\ % Newline to stack the figures vertically

    \vspace{0.25cm} % Adjust spacing

    % Second subfigure (Image)
    \subfloat[Sensitive Information Disclosure Attack flow à la MITRE ATLAS\label{fig:info_disclosure_atlas-flow}]{
        \includegraphics[width=0.99\textwidth]{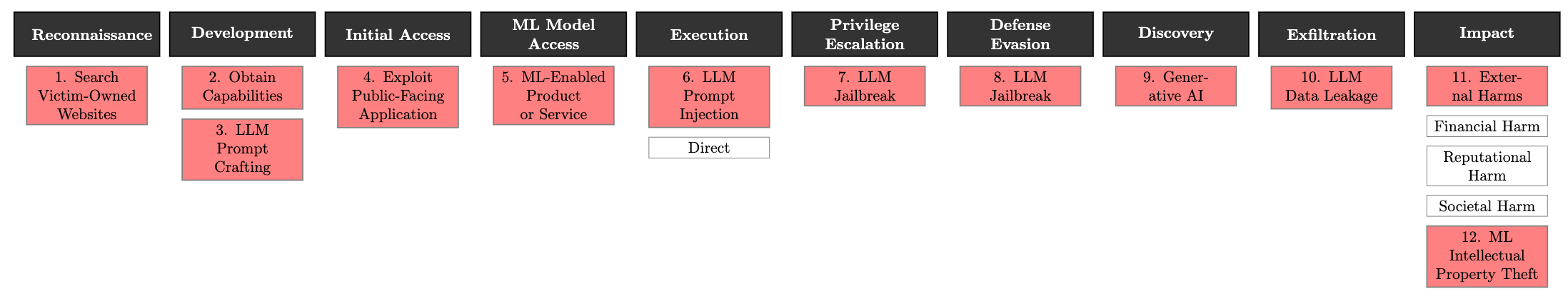}
    }

    \caption{Mapping Tactics, Techniques, and Procedures (TTPs) used in an example Sensitive Information Exfiltration campaign}
\end{figure}

Figure \ref{fig:info_leak} illustrates how adversaries craft poisoned retrieval queries and inject malicious tokens into the generation process, while Figure \ref{fig:info_disclosure_atlas-flow} integrates these steps to depict the end‑to‑end attack flow. 

These attacks target either data privacy, exposing training data, or model privacy. revealing internal model configurations, like system prompts. In RAG systems, the tight coupling between retrieval and generation increases the risk of cascading leaks across components.

\subsubsection{Threat Model II: RAG System Poisoning}

This threat model focuses on adversarial attempts to compromise the integrity and reliability of RAG systems by introducing malicious inputs or manipulating model parameters. Poisoning attacks are designed to degrade performance, embed hidden behaviors, or enable persistent leakage of sensitive information \cite{MITREATLAS_AMLT0018000,xue2024badrag, owasp_llm04}. In RAG systems, poisoning can occur across multiple stages of the AI pipeline, including:

\begin{itemize}
    \item \textbf{Training and Fine-Tuning:} Adversaries inject corrupted data into the initial training or fine-tuning process, compromising model weights and behaviors \cite{MITREATLAS_AMLT0018000}.
    
    \item \textbf{Document Ingestion:} Poisoned documents are inserted into ingestion pipelines, polluting the retrieval corpus and influencing downstream responses \cite{xue2024badrag}.
    
    \item \textbf{Retrieval and Indexing:} Attackers manipulate retrieval datasets or vector stores to embed adversarial payloads or mislead context retrieval \cite{MITREATLAS_AMLT0018000, OWASP_LLM08_VectorEmbedding}.
    
    \item \textbf{Prompt Engineering and System Prompts:} Maliciously crafted prompts or poisoned system instructions are introduced to destabilize output or bypass controls \cite{OWASP_LLM01_PromptInjection}.
    
    \item \textbf{Downstream Applications:} Third-party tools or integrated apps relying on model output become vectors for supply chain risk. \cite{OWASP_LLM03_SupplyChain}
\end{itemize}

\begin{figure}[ht]
    \centering

    % First subfigure
    \subfloat[RAG Architecture labeled with Tactics, Techniques, and Procedures (TTPs) for RAG Poisoning Attack via External Threat\label{fig:poison_rag_a}]{
        \includegraphics[width=0.94\linewidth]{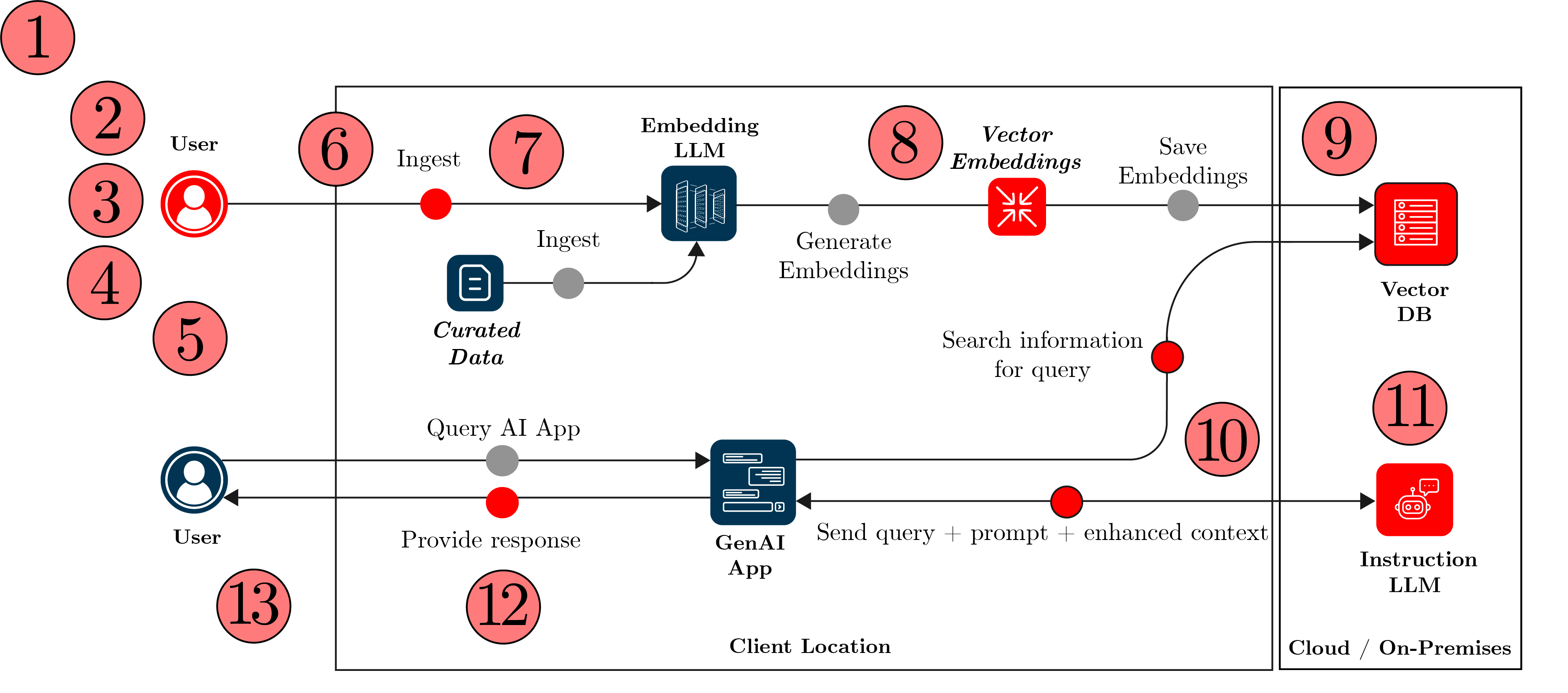}
    }\\

    \vspace{0.25cm} % Adjust spacing

    % Second subfigure
    \subfloat[RAG Architecture labeled with Tactics, Techniques, and Procedures (TTPs) for RAG Poisoning Attack via Insider Threat or Unwitting Insider\label{fig:poison_rag_b}]{
        \includegraphics[width=0.94\linewidth]{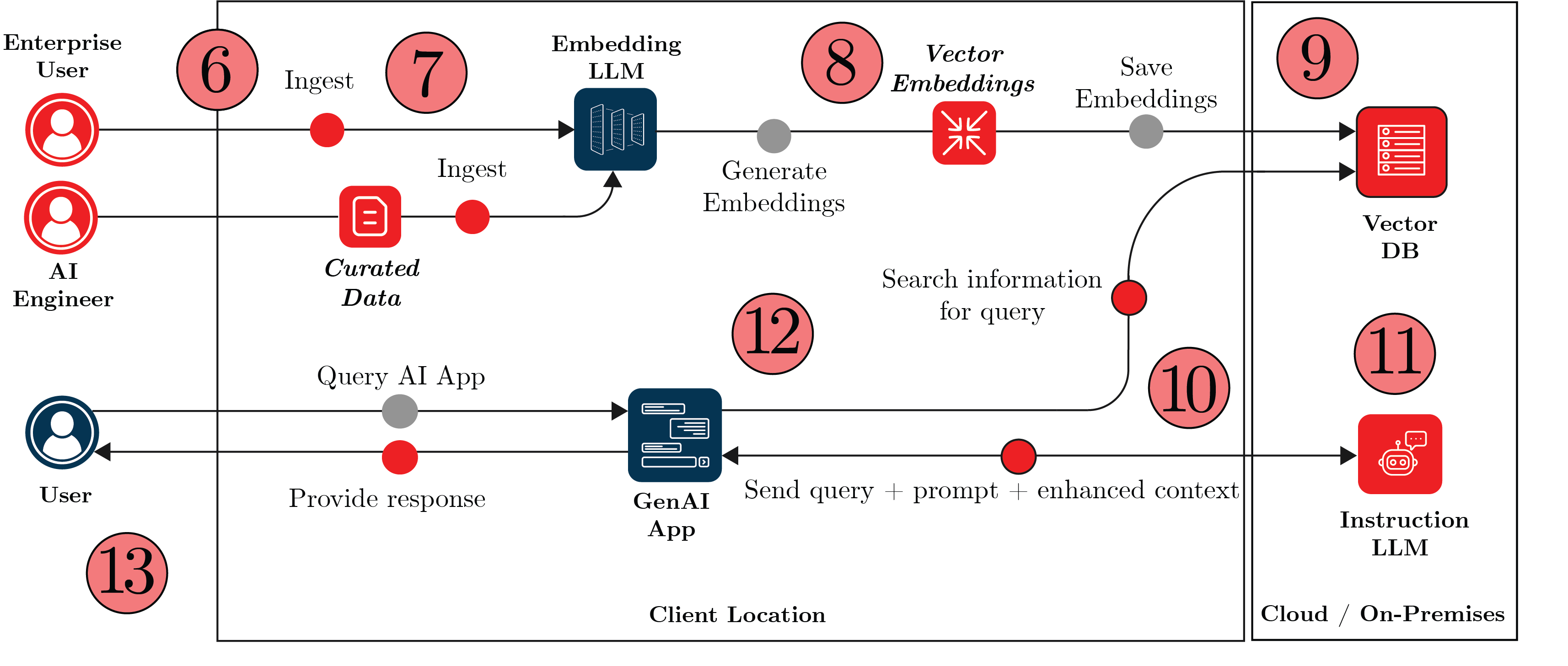}
    }\\

    \vspace{0.25cm} % Adjust spacing

    % Third subfigure
    \subfloat[RAG Poisoning Attack Flow à la MITRE ATLAS\label{fig:poison-rag-attack-flow}]{
        \includegraphics[width=0.99\linewidth]{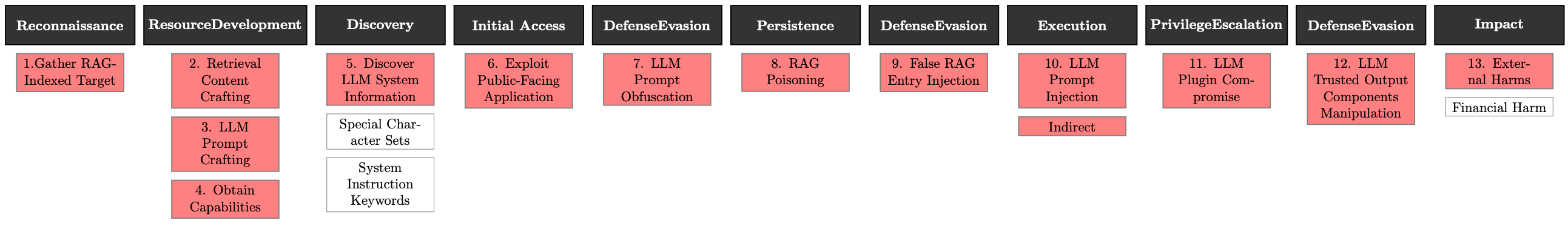}
    }

    \caption{Mapping Tactics, Techniques, and Procedures (TTPs) used in an example RAG Poisoning campaign}
    \label{fig:rag_poisoning_threat-model}
\end{figure}

\paragraph{External vs. Insider Threat Paths}

Figure \ref{fig:rag_poisoning_threat-model} illustrates two distinct RAG poisoning attack paths: (\ref{fig:poison_rag_a}) an external threat actor and (\ref{fig:poison_rag_b}) an insider threat or unwitting insider. Figure~\ref{fig:poison_rag_a} shows how an external adversary, operating outside the organizational boundary, targets public-facing ingestion points, typically exploiting insecure document upload interfaces or submitting poisoned data through legitimate channels. This actor must overcome perimeter defenses to inject malicious content, relying on open attack surfaces and indirect access to the retrieval pipeline.

In contrast, Figure~\ref{fig:poison_rag_b} maps the attack path of an insider threat or an unwitting insider. Here, the adversary operates from within the trusted environment, often as an employee or contractor with direct access to ingestion processes or document repositories. This positioning allows the threat actor to bypass certain external security controls, making it easier to insert poisoned data or manipulate vector embeddings with fewer immediate barriers. In some cases, well-meaning users may unintentionally contribute to poisoning by uploading unvetted or compromised documents without malicious intent.

These two pathways underscore a critical security takeaway: while external attacks tend to focus on exploiting exposed APIs and interfaces, insider threats leverage trusted roles and access rights, often requiring different detection and mitigation strategies. As a result, effective RAG defense must address both perimeter-focused and internal governance weaknesses, embedding continuous validation, monitoring, and access controls throughout the entire data lifecycle.

A key distinction is that insider threats, whether malicious or unwitting, fundamentally accelerate the attack process. Unlike external actors, who must progress through initial stages such as reconnaissance, capability development, and system discovery (steps 1–5 in Figure \ref{fig:poison-rag-attack-flow}), insiders are already positioned past these hurdles. They can move directly to poisoning activities (step 6 onward), drastically reducing the time-to-impact and the effort required to compromise the system. This acceleration makes insider-driven poisoning both faster and potentially harder to detect unless strong internal safeguards are in place.

\subsection{Assess and Identify Risks}
\label{sec:risk_quantification}

\begin{figure}[ht]
    \centering
    \includegraphics[width=0.85\linewidth]{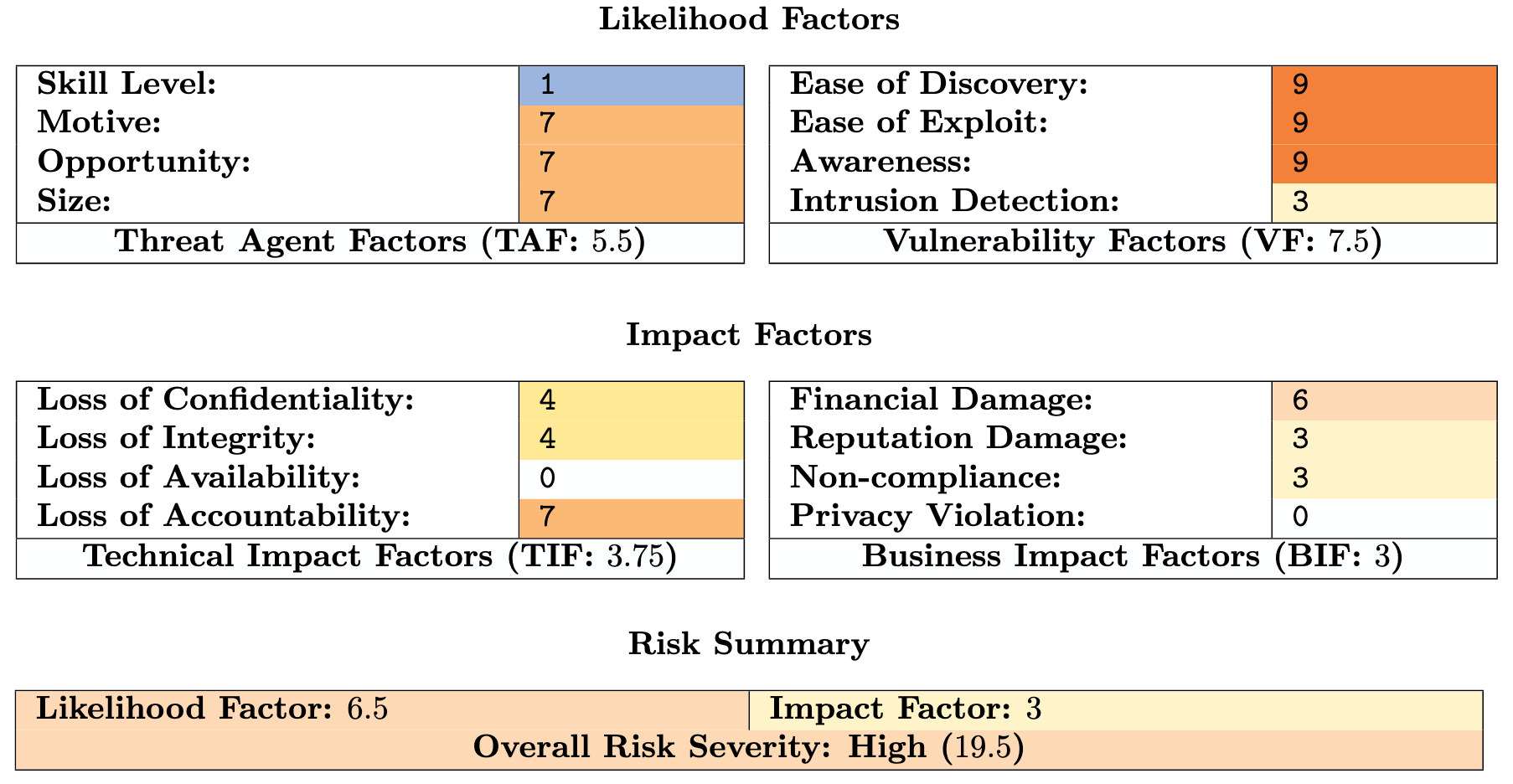}
    \caption{Inherent Risk Scoring for Sensitive Information Disclosure}
    \label{fig:rag_info_disclosure_inherent}
\end{figure}

The impacts of sensitive information disclosure include privacy violations, regulatory non-compliance, legal risk, reputational damage, and erosion of stakeholder trust. These risks are documented in OWASP Top 10 for LLM Applications~\cite{owasp:LLM02}, and align with multiple adversarial techniques from the MITRE ATLAS framework~\cite{mitre:atlas}. Figure \ref{fig:rag_info_disclosure_inherent} shows the inherent risk assessment for sensitive information disclosure, presenting the likelihood and impact scores for each risk vector before any mitigations are applied.

\begin{figure}[ht]
    \centering
    \includegraphics[width=0.85\linewidth]{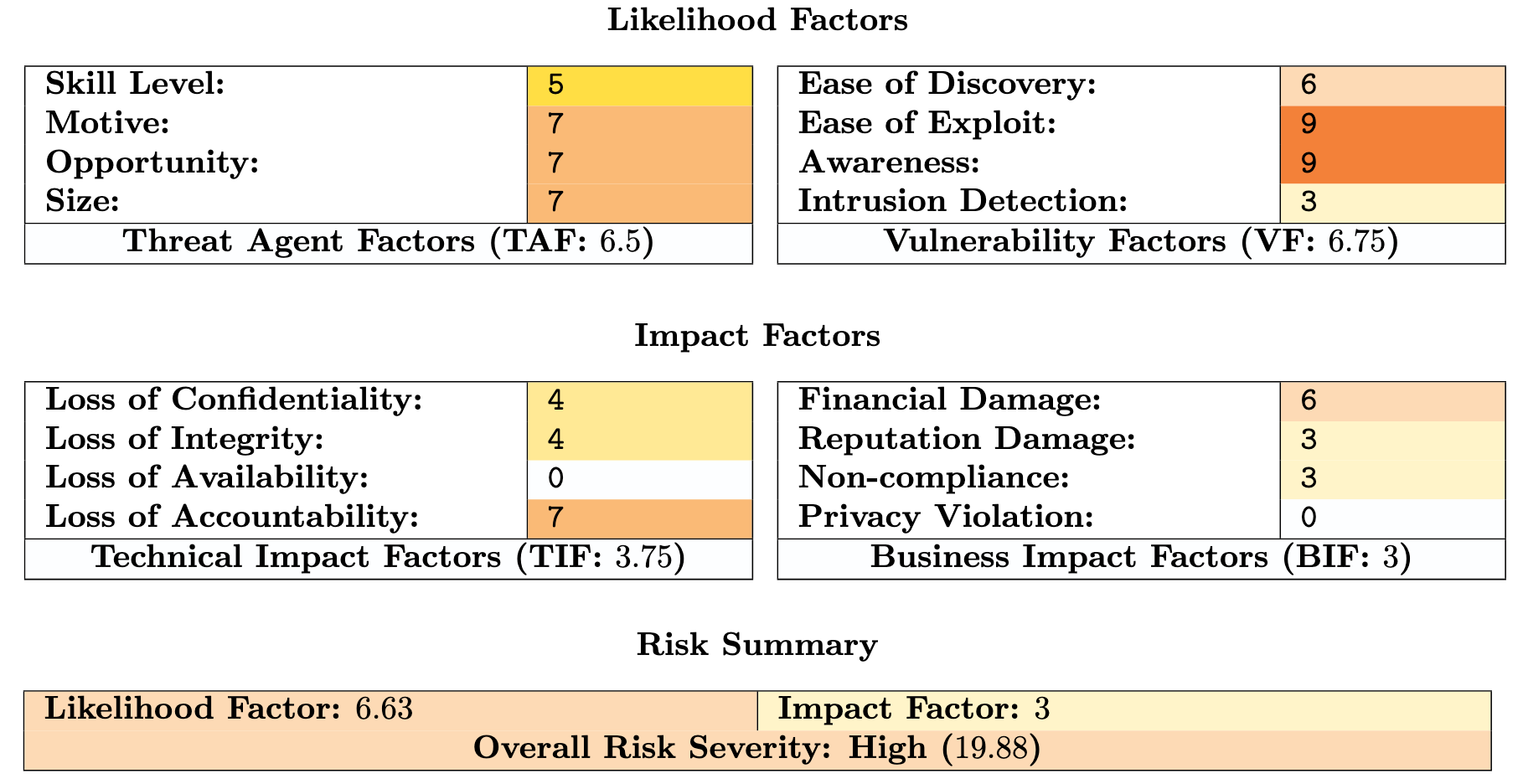}
    \caption{Inherent Risk Scoring for RAG Poisoning}
    \label{fig:rag_poison_risk}
\end{figure}

The consequences of poisoning attacks include biased or unreliable outputs, covert activation of hidden behaviors (such as backdoors), degraded system accuracy, and violations of policy or compliance mandates. These issues can remain undetected for extended periods, undermining trust in the system and leading to significant operational, reputational, and legal risks. Key assets at risk include model weights, vector embeddings, retrieval datasets, and the integrity of generated responses. We summarize the inherent risk factors for RAG system poisoning, including threat agent, vulnerability, and impact components, in the risk scoring model presented in Figure \ref{fig:rag_poison_risk}.

\clearpage

\subsection{Implement Controls and Validate}
\label{sec:risk_controls} 
Because these risks are manifested from system weaknesses, vulnerabilities in RAG systems cannot be remedied through a single patch or isolated update. Unlike traditional software flaws that may be resolved with targeted fixes, the risks in RAG systems span multiple layers from input handling to data governance and lifecycle management. This inherent complexity, combined with evolving attack techniques, necessitates a comprehensive mitigation strategy. Here we explore applicable risk controls for RAG systems. We show residual risk and estimated control efficacy in Figures \ref{fig:residual_SID}, \ref{fig:residual_rag_poison} respectively.

\subsubsection{Input Validation and Sanitization}
Input validation and sanitization involves enforcing rigorous checks on all incoming prompts and queries to identify and block malicious or malformed inputs. Clearly defined acceptable formats and character sets should be established, while known malicious patterns are systematically filtered out. Automated filters, leveraging either machine-learning techniques or rule-based systems, are implemented to proactively flag suspicious tokens, sequences, or prompts prior to processing. Effective controls here drive down \textit{Ease of Exploit, Opportunity,} and \textit{Loss of Integrity factors}.

\subsubsection{Adversarial Training and Testing}
Adversarial Training and Testing requires the generation of synthetic attack scenarios, such as deliberate prompt injection attempts, to continuously assess the model's resilience. Incorporating adversarially perturbed data during fine-tuning hardens the model against real-world threats. Additionally, penetration testing, often conducted by a red team, provides practical insights into hidden vulnerabilities in data pipelines and model interfaces. These practices can mitigate risks by increasing the \textit{Skill Level} required to exploit vulnerabilities (raising it from ``some technical skills" to ``security penetration skills"), reducing the \textit{Ease of Exploit} (moving it from ``easy" or ``automated tools available" toward ``difficult" or ``theoretical"), and limiting the \textit{Size} of potential threat agents (narrowing from ``anonymous internet users" to more specialized attackers).

\subsubsection{Real-Time Monitoring and Detection}
Real-time monitoring and detection involve continuous scrutiny of query patterns and outputs to identify anomalies, such as unusually frequent prompt requests or content irregularities. Detailed logging of user interactions facilitates forensic analysis and compliance reporting. Integrating these logs with Security Information and Event Management (SIEM) systems further enables automated alerts whenever suspicious activities are detected.

Real‑time monitoring and detection combine continuous telemetry collection with behaviour‑based analytics to spot abuse patterns the moment they emerge.  Every inference request, retrieval query, and model output is timestamped, tagged with a unique session ID, and pushed into a centralized log pipeline. 

Statistical baselining (bursty prompt‑submission, sudden spikes in retrieval‑error rates, or output entropies outside the learned norm.  These signals are enriched with user‑ and host‑level context and forwarded to a SIEM for correlation against other enterprise events. Automated response logic (Security Orchestration, Automation, and Response (SOAR) playbooks) can throttle, sandbox, or temporarily block the offending principal while human analysts investigate. From an OWASP risk‑rating lens, robust monitoring decreases \emph{Intrusion Detection} scores and indirectly raises the \emph{Skill Level} required for a successful exploit (because attackers must now evade layered anomaly detectors), it also shrinks \emph{Loss of Accountability} by ensuring high‑fidelity audit trails are available for forensic reconstruction.

\subsubsection{Data Governance and Curation} 
A cross‑functional data‑governance committee defines the approved data sources and rule sets, reviews exceptions, and audits compliance on a regular cadence.  Under this framework, every dataset that can influence retrieval or model fine‑tuning is subjected to strict hygiene controls.  External corpora first pass through a secure staging zone where schema validation, MIME‑type whitelisting, and multi‑engine malware scanning run in parallel.  Content then undergoes automated and/or manual review to strip sensitive information such as Personally Identifiable Information (PII), profanity, and policy‑violating text, followed by fact‑consistency checks against trusted references.  Only documents and artifacts that clear every gate are version‑pinned, cryptographically signed, and stored immutably; provenance metadata (origin URL, hash, ingestion timestamp) is written to a tamper‑evident ledger to preserve chain‑of‑custody.  Role‑based access controls and encrypted transfer protocols prevent unauthorized edits and eavesdropping.  Collectively, these controls slash the \emph{Opportunity} available to threat actors, raise the \emph{Ease of Exploit} bar for data‑poisoning attacks, and curb both \emph{Loss of Integrity} and downstream \emph{Reputation Damage} or \emph{Privacy Violations} by ensuring that only vetted, traceable knowledge reaches production.

\subsubsection{AI Lifecycle Management and Machine Learning Operations (MLOps)}
Lifecycle management ensures that AI systems remain secure, reliable, and effective from initial development through sustained operation. A robust, assured deployment pipeline, anchored in continuous integration and continuous deployment (CI/CD) practices, is critical. The CRISP-ML(Q) framework\cite{studer2021towards} provides an excellent foundation for structuring this full lifecycle, emphasizing traceability, transparency, and quality assurance at every stage.

% -------------------------------------------------
\subparagraph{Phase 1. Business \& Data Understanding}% 
Document requirements, success criteria, data‑handling rules, and regulatory limits before any coding starts, so that every later phase can be checked against the same standards. These actions clarify acceptable use up‑front, which lowers \emph{Opportunity} for unnecessary, unauthorized, or out‑of‑scope data collection and reduces downstream \emph{Reputation Damage} if the project is questioned.

% -------------------------------------------------
\subparagraph{Phase 2. Data Engineering}%
Data engineering processes establish immutable provenance and traceability for data from initial collection through final storage. Data scientists and engineers carefully curate datasets to ensure effective model training, reduce biases, and maintain data quality standards. Each stage produces a \emph{Data Card}, documenting data origin, transformations, and integrity checks, directly informing the AI System Bill of Materials (AI BOM). Measures here significantly reduce adversarial \emph{Opportunity} for data tampering and raise the \emph{Ease of Exploit} threshold for dependency-based attacks.

\subparagraph{Phase 3. Model Engineering}
With respect to AI safety and security, static analysis and test suites are used to identify vulnerabilities, assess common weaknesses, evaluate bias and limitations in this phase. Performance benchmarks are established to enable ongoing monitoring for drift, anomalies, and degraded inference quality. Models are versioned and frozen upon release to ensure reproducibility and auditability, with all key details captured in a \emph{Model Card}, which is integrated into the AI System Bill of Materials (AI BOM). Early defect detection in this phase shifts \emph{Ease of Exploit} from “easy” toward “difficult,” maintaining a low potential for \emph{Loss of Integrity}.

\subparagraph{Phase 4. Model Evaluation}
After training, models undergo rigorous evaluation to validate performance, robustness, and suitability for deployment. Testing is performed on held-out datasets to confirm generalization, while additional stress tests using noisy, adversarial, or edge-case data assess the model’s resilience. Evaluation criteria are aligned with business and regulatory requirements, covering accuracy, fairness, explainability, and resource efficiency.
Where applicable, explainability tools are applied to ensure transparency and foster trust, especially in high-stakes domains. Evaluation results, along with reproducibility checks (e.g., multiple random seeds), are documented thoroughly in the \emph{Model Card} to support oversight and compliance. Careful evaluation at this stage helps catch hidden defects, further reducing \emph{Loss of Integrity} potential and increasing system resilience before production deployment.

\subparagraph{Phase 5. Deployment}
New models are deployed gradually, initially serving a small test group. Key health signals—such as performance, error rates, and latency—are closely monitored; if metrics remain within acceptable thresholds, rollout proceeds to broader audiences. If issues arise, rapid rollback mechanisms restore the previous stable version with minimal delay. Progressive rollout strategies limit the blast radius of failures, reducing \emph{Loss of Availability} and minimizing potential \emph{Financial Damage} or \emph{Reputation Damage} from faulty releases.

\subparagraph{Phase 6. Monitoring \& Maintenance}
Once deployed, ML models require continuous monitoring to ensure stable, reliable performance in real-world conditions. A primary risk is \emph{model staleness}, where accuracy and effectiveness degrade as the model encounters new or shifting data patterns. Performance can also be impacted by changes in hardware or software environments.

Best practice follows the \emph{Continued Model Evaluation} pattern: models are routinely evaluated against fresh data to detect drift, anomalies, or performance degradation early.\cite{lakshmanan2020machine}. Monitoring insights drive decisions on retraining, model replacement, or process adjustments to maintain alignment with business objectives. All configuration changes and retraining cycles are tracked in version control and require peer approval, ensuring transparency and governance. Continuous evaluation and disciplined maintenance keep the \emph{Ease of Exploit} high for would-be attackers and minimize the window for successful attacks, sustaining a low overall \emph{Likelihood} of compromise.

\subsubsection{Incident Response and Recovery}
Incident Response is a critical control that provides a structured, rehearsed plan for detecting, containing, and recovering from security incidents. In the context of RAG systems, where data and model integrity are paramount, the Incident Response Plan (IRP) ensures that any breach, whether through data poisoning, model tampering, or infrastructure compromise, is rapidly addressed to minimize damage.
The Incident Response Plan outlines specific procedures: immediate isolation of compromised components or data sources, invocation of automated and manual playbooks, and coordinated response across security, IT, and AI/ML teams. Maintaining secure, up to date backups of critical models, datasets, and configuration files is essential to restore operations quickly and limit downtime.
Beyond technical response, the Incident Response Plan defines roles and responsibilities, escalation paths, legal and compliance notification requirements, and communication strategies to manage stakeholder expectations and preserve organizational trust.
Post incident reviews (root cause analysis and lessons learned) are mandatory steps, feeding improvements back into the overall security posture. This process strengthens resilience by identifying control gaps, fine tuning detection capabilities, and iterating on response readiness.
As a formal risk control, Incident Response and Recovery mitigate potential \emph{Loss of Availability}, \emph{Integrity}, and \emph{Reputation Damage} by ensuring that even when preventive defenses fail, impact is contained and recovery is swift, preserving business continuity and regulatory compliance.

\subsubsection{Control Prioritization: AI Security Pyramid of Pain Analysis}
\label{sec:risk_control_prioritization}

\label{sec:pyramid_of_pain_analysis}

The AI Security Pyramid of Pain prioritizes controls based on how robust they are and the degree of disruption, or ``pain", they inflict on adversaries. At the lower levels of the pyramid, data integrity measures are relatively easy for an adversary to adapt to; these issues can typically be patched with targeted fixes, and adversaries can quickly modify their tactics. In contrast, higher up in the pyramid lie threats that are intertwined with the adversary's tactics, techniques, and procedures (TTPs). Mitigations at these levels, such as robust data governance and comprehensive adversarial training, force adversaries to fundamentally alter their operational approach, which in turn imposes significant costs and delays on their efforts.

By aligning risk mitigation strategies with the upper tiers of the pyramid, organizations can impose a greater operational burden on adversaries. Controls that target the intrinsic aspects of data poisoning or adversarial query manipulation, for example, require adversaries to redesign their entire attack methodology rather than simply circumvent a superficial patch. This layered defense strategy not only strengthens the overall security posture of RAG systems but also maximizes the adversary's difficulty in resourcing and sustaining effective attacks.

Based on the AI Security Pyramid of Pain, the Table \ref{tab:PoP_Map} ranks the controls by the degree of disruption they inflict on adversaries. Controls at the top force adversaries to fundamentally change their tactics, while those lower in the ranking are more easily bypassed or reactive in nature.

Many of the implemented controls and CRISP-ML lifecycle phases span multiple tiers of the AI Security Pyramid of Pain. Upper-tier controls impose significant disruption, forcing adversaries to rethink their tactics and methodologies, while lower-tier controls provide essential protections and maintain overall system resilience.

Table~\ref{tab:PoP_Map} provides a structured mapping of these controls and CRISP-ML phases to the corresponding Pyramid layers, highlighting how each mitigation contributes across different tiers of defense.

\vspace{0.25cm}

\begin{table}[h]
\caption{Multi-Level Mapping of Controls and CRISP-ML Phases to the AI Security Pyramid of Pain}
\label{tab:PoP_Map}
\vspace{0.3cm}
\centering
\begin{tabular}{|p{0.2\linewidth}|p{0.75\linewidth}|}
\hline
\textbf{Pyramid Layer} & \textbf{Mapped Controls and Phases} \\ \hline
\textbf{TTPs} &
- Adversarial Training and Testing \newline
- CRISP-ML Phase 3: Hardening through static analysis and version control \\ 
\hline
\textbf{Data Provenance} &
- Data Governance and Curation \newline
- Data Engineering (Data Cards and AI BOM) \newline
- CRISP-ML Phase 2: Data Engineering (provenance and traceability enforcement) \newline
- Lifecycle Management (embedding provenance into CI/CD pipelines) \\ \hline
\textbf{Adversarial Inputs} &
- Input Validation and Sanitization \newline
- CRISP-ML Phase 1: Business and Data Understanding\newline
- CRISP-ML Phase 3: Model Engineering (prompt injection testing) \\ \hline
\textbf{Adversarial Tools} &
- Integration of Red Teaming Tools \newline
- CRISP-ML Phase 4: Model Evaluation (leverage adversarial tools for testing) \newline
- Real-Time Monitoring and Detection (detection of automated adversarial scripts) \\ \hline
\textbf{AI System Performance} &
- Model Evaluation (stress testing, explainability audits) \newline
- Deployment Monitoring and Maintenance (benchmarking and drift detection) \newline
- CRISP-ML Phase 5: Deployment (progressive rollout strategies) \newline
- CRISP-ML Phase 6: Monitoring and Maintenance (continuous evaluation) \\ \hline
\textbf{Data Integrity} &
- Incident Response and Recovery (containment and restoration) \newline
- Incident Response (root cause analysis and remediation) \\ \hline
\end{tabular}
\end{table}

\clearpage

\subsubsection{Validation and Residual Risk Measurement}
\label{sec:validation_residual_risk}

In this section, we reassess risks, following control implementation, to understand the remaining residual risk and its alignment with acceptable thresholds. For each control, we assess its impact by analyzing how it limits the adversary’s capability or opportunity. For example, examining how input validation blocks specific injection vectors. The resulting risk profile reflects the reassessed likelihood and impact, providing a clear indication of the remaining exposure after mitigation measures are applied. Following the application of mitigating controls, the numerical risk profiles for both primary threat models showed substantial reductions. 

\begin{figure}[!ht]
    \centering
    \includegraphics[width=0.85\linewidth]{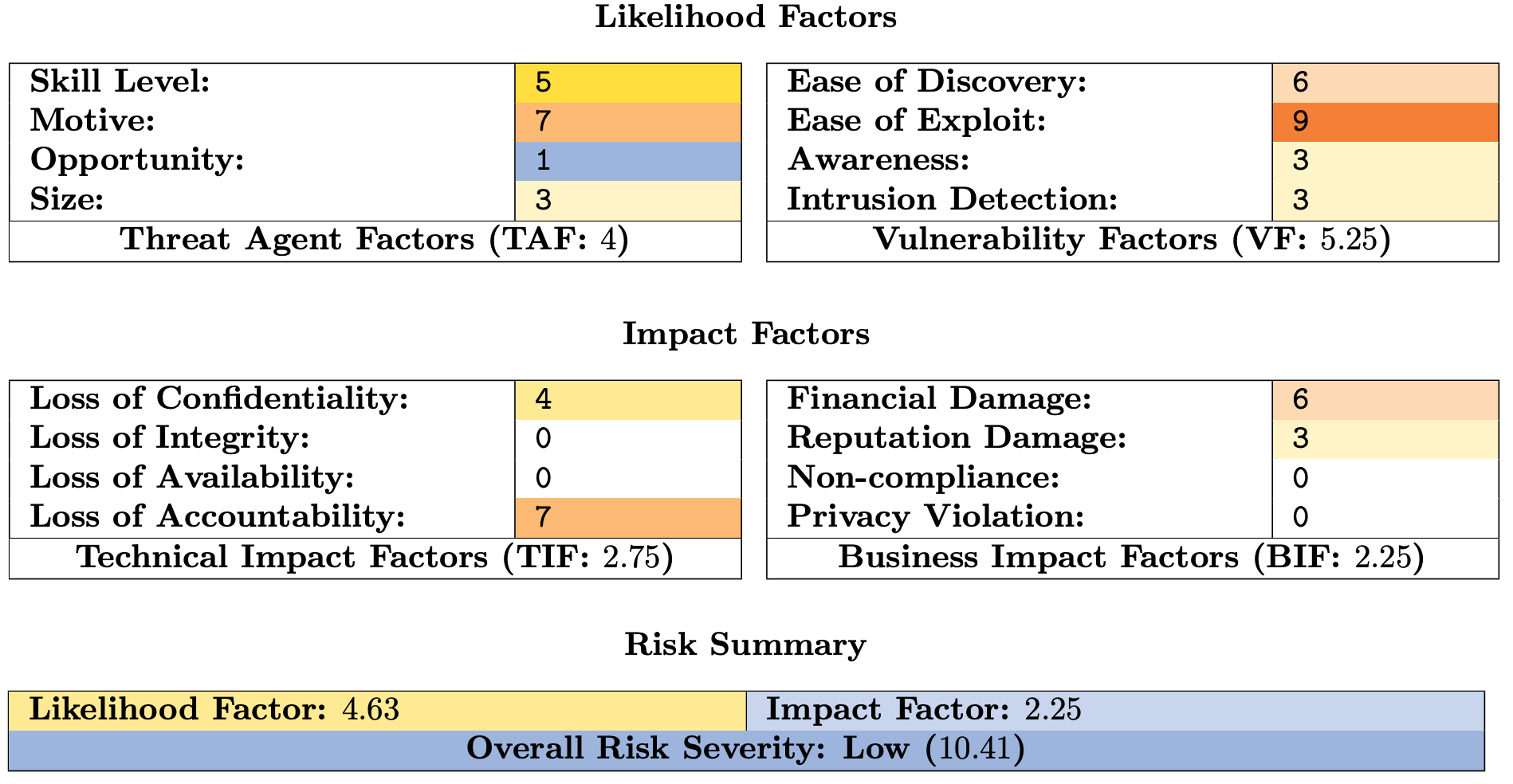}
    \vspace{0.5cm}
    \caption{Residual Risk Scoring for Sensitive Information Disclosure}
    \label{fig:residual_SID}
\end{figure}

For Sensitive Information Disclosure (Threat Model I), we observed a meaningful reduction in residual risk, as detailed in Figure~\ref{fig:residual_SID}. The Overall Risk Severity decreased from High (19.5), to Low (10.41); this was driven by its Likelihood Factor dropping from 6.5 to 4.63 and its Impact Factor dropping to 2.25 from 3. 

\begin{figure}[!ht]
    \centering
    \includegraphics[width=0.85\linewidth]{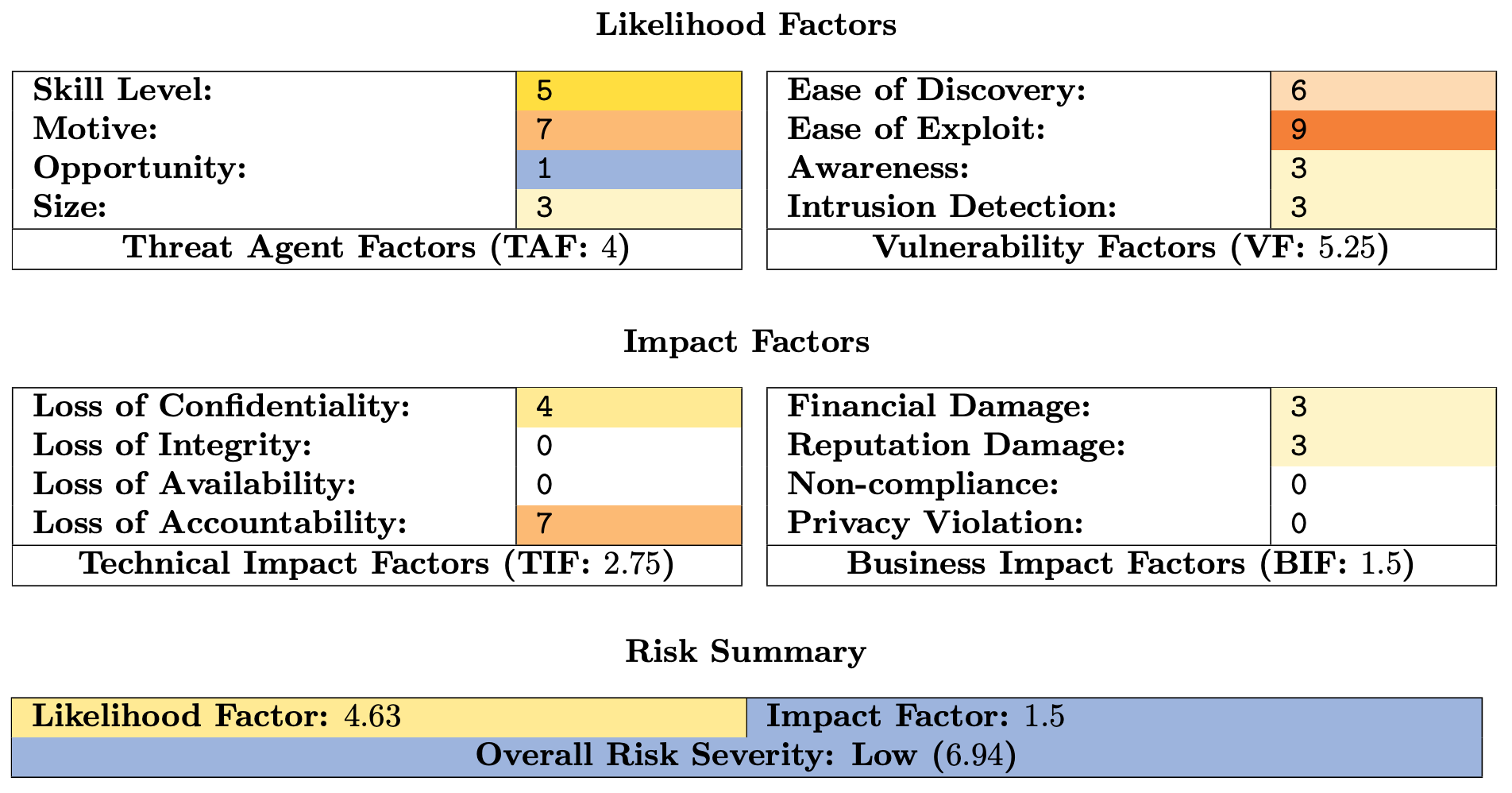}
    \vspace{0.5cm}
    \caption{Residual Risk Scoring for RAG Poisoning}
    \label{fig:residual_rag_poison}
\end{figure}

Similarly, for RAG System Poisoning (Threat Model II), the Overall Risk Severity was reduced from High (19.88), to Low (6.94). The application of strict data governance and lifecycle controls significantly curtailed the adversary’s ability to persist within the retrieval pipeline, as illustrated in Figure~\ref{fig:residual_rag_poison}. This results in a decrease in overall Likelihood Factor from 6.63 to 4.63 and a significant reduction in Impact Factor from 3 to 1.5. These quantitative improvements underscore the positive impact of the implemented mitigation strategies. While poisoning threats cannot be fully eradicated, the residual risk now sits within a low-risk bracket, supported by enhanced detection and containment capabilities.

\goodbreak

\goodbreak

%\begin{center}
%  \includegraphics[width=0.99\textwidth]{figures/risk_calculation_rag_poisoning_insider.png}
%\end{center}

\section{Discussion and Future Work}
\label{sec:discussion_future_work}

Our analysis demonstrates that a multi-layered mitigation strategy, aligned with the AI Security Pyramid of Pain, can meaningfully reduce the inherent risks associated with RAG systems. By addressing key adversarial vectors, such as prompt injection, data poisoning, and adversarial query manipulation, we achieved significant risk drawdown across the attack surface. The integration of rigorous input validation, adversarial training, proactive data governance, and continuous monitoring proved particularly effective in driving down both the likelihood and impact of adversarial exploits.

Despite these successes, certain residual risks remain unavoidable due to the dynamic and evolving nature of adversarial tactics. For example, sophisticated insider threats or advanced supply chain compromises will continue to challenge even well-defended RAG architectures. Additionally, while our mitigation strategy effectively reduced opportunity and ease of exploit, the persistence of latent weaknesses and vulnerabilities in large, distributed systems requires ongoing vigilance and adaptive countermeasures.

Future work should focus on deepening system resilience and advancing operational assurance. Empirical validation through red teaming exercises and live adversarial testing is critical to verify the robustness of the proposed mitigation strategies under real-world conditions. By coupling formal risk assessments with hands-on stress testing, organizations can more confidently operationalize RAG systems in sensitive environments while maintaining a strong security posture.

\section{Conclusion}
\label{sec:conclusion} 
As adversarial tactics evolve, securing RAG systems demands a comprehensive strategy that spans all layers of the AI Security Pyramid of Pain. By integrating robust defenses against prompt injection, data poisoning, and adversarial query manipulation, organizations can better protect the integrity and reliability of their systems. Continuous monitoring, improved input sanitization, and proactive data curation are essential to mitigate these risks and support the ongoing adoption of RAG technology in critical domains
This structured approach enables organizations to effectively mitigate RAG-specific threats while ensuring the reliability, trustworthiness, and compliance of their AI-driven applications.

\section{Acknowledgments}
We would like to thank our colleague Dr. Mike Tan for the helpful conversations and insights that informed this work. His perspective contributed to our thinking during the development of the material, and we appreciate his input during the research process.

\bibliography{references}

\begin{thebibliography}{10}

\bibitem{lewis2020retrieval}
Lewis, P., Perez, E., Piktus, A., Petroni, F., Karpukhin, V., Goyal, N., K{\"u}ttler, H., Lewis, M., Yih, W.-t., Rockt{\"a}schel, T., et~al., ``Retrieval-augmented generation for knowledge-intensive nlp tasks,'' {\em Advances in neural information processing systems}~{\bf 33},  9459--9474 (2020).

\bibitem{menlo2024}
Tully, T., Redfern, J., and Xiao, D., ``2024: The state of generative ai in the enterprise.'' Menlo Ventures Blog (Nov 2024).
\newblock Industry survey: 51\% enterprise adoption of RAG, up from 31\% in 2023.

\bibitem{babeanu2025}
Babeanu, ``Is your rag a security risk?.'' RSA Conference Blog (Feb 2025).
\newblock Industry analysis of GenAI risks including prompt injection and data exposure.

\bibitem{xue2024badrag}
Xue, J., Zheng, M., Hu, Y., Liu, F., Chen, X., and Lou, Q., ``Badrag: Identifying vulnerabilities in retrieval augmented generation of large language models,'' {\em arXiv preprint arXiv:2406.00083}  (2024).

\bibitem{rieder2021mapping}
Rieder, G., Simon, J., and Wong, P.-H., ``Mapping the stony road toward trustworthy {AI}: Expectations, problems, conundrums,'' in [{\em Machines We Trust: Perspectives on Dependable {AI}}{\nolinebreak\hspace{0.1em}]},  Pelillo, M. and Scantamburlo, T., eds.,  27--40, The MIT Press, Cambridge, MA (2021).

\bibitem{Dahj2022MasteringCI}
Dahj, J. N.~M.,  [{\em Mastering Cyber Intelligence: Gain comprehensive knowledge and skills to conduct threat intelligence for effective system defense}{\nolinebreak\hspace{0.1em}]}, Packt Publishing (2022).

\bibitem{ai2023artificial}
AI, N., ``Artificial intelligence risk management framework (ai rmf 1.0),'' {\em URL: https://nvlpubs. nist. gov/nistpubs/ai/nist. ai} ,  100--1 (2023).

\bibitem{Shostack2014}
Shostack, A.,  [{\em Threat Modeling: Designing for Security}{\nolinebreak\hspace{0.1em}]}, Wiley (February 2014).

\bibitem{ward2024ai}
Ward, C.~M., Harguess, J., Tao, J., Christman, D., Tan, M., Spicer, P., and Cranium, A., ``The ai security pyramid of pain,'' in [{\em Proc. of SPIE Vol}{\nolinebreak\hspace{0.1em}]},   {\bf 13054},  1305408--1 (2024).

\bibitem{vaswani2017attention}
Vaswani, A., Shazeer, N., Parmar, N., Uszkoreit, J., Jones, L., Gomez, A.~N., Kaiser, {\L}., and Polosukhin, I., ``Attention is all you need,'' {\em Advances in neural information processing systems}~{\bf 30} (2017).

\bibitem{zhao2023survey}
Zhao, W.~X., Zhou, K., Li, J., Tang, T., Wang, X., Hou, Y., Min, Y., Zhang, B., Zhang, J., Dong, Z., et~al., ``A survey of large language models,'' {\em arXiv preprint arXiv:2303.18223}~{\bf 1}(2) (2023).

\bibitem{mitre_cwe_overview}
{The MITRE Corporation}, ``Common weakness enumeration (cwe).'' \url{https://cwe.mitre.org/}.
\newblock Accessed: 2025-03-17.

\bibitem{cwe_ai_wg}
{The MITRE Corporation}, ``Cwe community working groups \& special interest groups.'' \url{https://cwe.mitre.org/community/working_groups.html}.
\newblock Accessed: 2025-03-17.

\bibitem{mitre2006cwe}
{The MITRE Corporation}, ``Common weakness enumeration (cwe).'' \url{https://cwe.mitre.org/} (2006).
\newblock Accessed: 2025-05-06.

\bibitem{martin2008common}
Martin, R.~A. and Barnum, S., ``Common weakness enumeration (cwe) status update,'' {\em ACM SIGAda Ada Letters}~{\bf 28}(1),  88--91 (2008).

\bibitem{CWE-20}
{The MITRE Corporation}, ``{CWE-20: Improper Input Validation}.'' \url{https://cwe.mitre.org/data/definitions/20.html} (February 2024).
\newblock Part of the Common Weakness Enumeration (CWE). Sponsored by the U.S. Department of Homeland Security (DHS) Cybersecurity and Infrastructure Security Agency (CISA). Accessed: May 6, 2025.

\bibitem{cve-2023-5423}
{CVE Program}, ``{CVE-2023-5423: SQL Injection in SourceCodester Online Pizza Ordering System 1.0}.'' \url{https://cve.org/CVERecord?id=CVE-2023-5423} (2023).
\newblock Accessed: 2025-05-07.

\bibitem{CWE-1039}
{The MITRE Corporation}, ``{CWE-1039: Automated Code Generation Based on Stale Schemas or Specifications}.'' \url{https://cwe.mitre.org/data/definitions/1039.html} (April 2023).
\newblock Part of the Common Weakness Enumeration (CWE). Sponsored by the U.S. Department of Homeland Security (DHS) Cybersecurity and Infrastructure Security Agency (CISA). Accessed: May 6, 2025.

\bibitem{CWE-1426}
{The MITRE Corporation}, ``{CWE-1426: Unintended Exposure of Sensitive Information in Generated Code}.'' \url{https://cwe.mitre.org/data/definitions/1426.html} (February 2024).
\newblock Part of the Common Weakness Enumeration (CWE). Sponsored by the U.S. Department of Homeland Security (DHS) Cybersecurity and Infrastructure Security Agency (CISA). Accessed: May 6, 2025.

\bibitem{microsoftThreatModeling}
Center, M. S.~E., ``Threat modeling.'' \url{https://www.microsoft.com/en-us/securityengineering/sdl/threatmodeling} (2022).
\newblock Accessed March 2025.

\bibitem{doyle2021vulnerability}
Doyle, M., Harguess, J., Manville, K., and Rodriguez, M., ``The vulnerability of uavs: An adversarial machine learning perspective,'' in [{\em Geospatial Informatics XI}{\nolinebreak\hspace{0.1em}]},   {\bf 11733},  81--92, SPIE (2021).

\bibitem{mitre:AML.T0024.000}
{MITRE ATLAS}, ``Infer training data membership (aml.t0024.000).'' \url{https://atlas.mitre.org/techniques/AML.T0024.000} (2023).

\bibitem{mitre:AML.T0024.001}
{MITRE ATLAS}, ``Invert ml model (aml.t0024.001).'' \url{https://atlas.mitre.org/techniques/AML.T0024.001} (2023).

\bibitem{mitre:AML.T0051.000}
{MITRE ATLAS}, ``Prompt leakage (aml.t0051.000).'' \url{https://atlas.mitre.org/techniques/AML.T0051.000} (2023).

\bibitem{OWASP_LLM07_PromptLeakage}
{OWASP Foundation}, ``Llm07:2025 system prompt leakage – owasp top 10 for llm applications.'' \url{https://genai.owasp.org/llmrisk/llm072025-system-prompt-leakage/} (2025).
\newblock Accessed: 2025-05-06.

\bibitem{mitre:AML.T0024.002}
{MITRE ATLAS}, ``Exploit vector embeddings (aml.t0024.002).'' \url{https://atlas.mitre.org/techniques/AML.T0024.002} (2023).

\bibitem{OWASP_LLM08_VectorEmbedding}
{OWASP Foundation}, ``Llm08:2025 vector and embedding weaknesses – owasp top 10 for llm applications.'' \url{https://genai.owasp.org/llmrisk/llm082025-vector-and-embedding-weaknesses/} (2025).
\newblock Accessed: 2025-05-06.

\bibitem{MITREATLAS_AMLT0018000}
{MITRE ATLAS}, ``Aml.t0018.000 – poison training data.'' \url{https://atlas.mitre.org/techniques/AML.T0018.000} (2023).
\newblock Accessed: 2025-05-06.

\bibitem{owasp_llm04}
Foundation, O., ``Llm04:2025 data and model poisoning,'' (2025).
\newblock Accessed: 2025-03-24.

\bibitem{OWASP_LLM01_PromptInjection}
{OWASP Foundation}, ``Llm01: Prompt injection – owasp top 10 for llm applications.'' \url{https://genai.owasp.org/llmrisk/llm01-prompt-injection/} (2025).
\newblock Accessed: 2025-05-06.

\bibitem{OWASP_LLM03_SupplyChain}
{OWASP Foundation}, ``Llm03:2025 supply chain vulnerabilities – owasp top 10 for llm applications.'' \url{https://genai.owasp.org/llmrisk/llm032025-supply-chain/} (2025).
\newblock Accessed: 2025-05-06.

\bibitem{owasp:LLM02}
{OWASP Foundation}, ``{OWASP Top 10 for LLM Applications – LLM02:2025 Sensitive Information Disclosure}.'' \url{https://owasp.org/www-project-top-10-for-large-language-model-applications/} (2025).

\bibitem{mitre:atlas}
{The MITRE Corporation}, ``{MITRE ATLAS Adversarial Threat Landscape for Artificial-Intelligence Systems}.'' \url{https://atlas.mitre.org} (2023).

\bibitem{studer2021towards}
Studer, S., Bui, T.~B., Drescher, C., Hanuschkin, A., Winkler, L., Peters, S., and M{\"u}ller, K.-R., ``Towards crisp-ml (q): a machine learning process model with quality assurance methodology,'' {\em Machine learning and knowledge extraction}~{\bf 3}(2),  392--413 (2021).

\bibitem{lakshmanan2020machine}
Lakshmanan, V., Robinson, S., and Munn, M.,  [{\em Machine Learning Design Patterns: Solutions to Common Challenges in Data Preparation, Model Building, and MLOps}{\nolinebreak\hspace{0.1em}]}, O'Reilly Media, Sebastopol, CA (2020).

\end{thebibliography}
\bibliographystyle{spiebib}

\end{document}